# Description of the cattle and small ruminants trade network in Senegal and implication for the surveillance of animal diseases

Mamadou Ciss[1§], Alessandra Giacomini[2,3,4§], Mame Nahé Diouf[1], Alexis Delabouglise[2,3], Asma Mesdour[2,3], Katherin Garcia Garcia[2,3], Facundo Munoz[2,3], Eric Cardinale[2,3], Mbargou Lo[5], Adji Marème Gaye[5], Mathioro Fall[5], Khady Ndiaye[5], Assane Guèye Fall[1], Catherine Cetre-Sossah[2,3], Andrea Apolloni[2,3]

[1] Institut Sénégalais de Recherches Agricoles/Laboratoire National de l'Elevage et de Recherches Vétérinaires BP 2057 Dakar-Hann, Sénégal
[2] CIRAD, UMR ASTRE, Montpellier, France
[3] CIRAD, UMR ASTRE, Univ Montpellier, INRAE, Montpellier, France
[4] Department of Biosciences, Swansea University, Swansea, SA2 8PP, UK
[5] Direction des Services Vétérinaires, Dakar, Sénégal
[§] M. C. and A. G. contributed equally

Corresponding author: Alessandra Giacomini; a.giacomini.2156511@swansea.ac.uk

## Abstract

Livestock mobility, particularly that of small and large ruminants, is one of the main pillars of production and trade in West Africa: livestock is moved around in search of better grazing or sold in markets for domestic consumption and for festival-related activities. These movements cover several thousand kilometers and have the capability of connecting the whole West African region thus facilitating the diffusion of many animal and zoonotic diseases. Several factors shape mobility patterns even in normal years and surveillance systems need to account for such changes. In this paper, we present a procedure based on temporal network theory to identify possible sentinel locations using two indicators: *vulnerability* (i.e. the probability of being reached by the disease) and *time of infection* (i.e. the time of first arrival of the disease). Using these indicators in our structural analysis of the changing network enabled us to identify a set of nodes that could be used in an early warning system.

As a case study we simulated the introduction of F.A.S.T. (Foot and Mouth Similar Transboundary) diseases in Senegal and used data taken from 2020 Sanitary certificates (LPS – *laissez-passer sanitaire*) issued by the Senegalese Veterinary Services to reconstruct the national mobility network. Our analysis showed that a static approach can significantly overestimate the speed and the extent of disease propagation, whereas temporal analysis revealed that the reachability and vulnerability of the different administrative departments (used as nodes of the mobility network) change over the course of the year. For this reason, several sets of sentinel nodes were identified in different periods of the year, underlining the role of temporality in shaping patterns of disease diffusion.

## 1. Introduction

The West African region includes the southern part of the bulge in the African continent and is crossed by the Sahel, a transitional strip between the Sahara Desert in the north and the Sudanic zone in the south (Bossard, 2009). The region is composed of 18 countries and is bounded in the north by Mauritania, Mali and Niger, in the east by Chad and Cameroon, in the south and west by the Atlantic Ocean. The region is characterized by different climates, and hence, different agro-ecological zones and different livestock farming systems (Missohou et al., 2016). Livestock farming (particularly cattle and small ruminants) is one of the most important economic activities in this area.

In West Africa, livestock mobility is an intrinsic component of livestock production and trade. The harsh environmental conditions, as well as the absence of the facilities required to slaughter animals and store meat, means livestock has to be mobile. To optimize the use of natural resources such as pasture and surface water, whose availability varies throughout the year, livestock farmers are forced to move their herds around: these movements occur all the year round (nomadism) or in specific periods (transhumance). Because of the lack of storage facilities and infrastructure, the majority of animals are sold alive at markets all year round. Most animals are concentrated in the northern part of West Africa, notably in Mali, Chad, Niger, and Mauritania, where the vast uninhabited areas are unsuitable for cropping but allow extensive livestock raising, and the animals are moved towards the greener southern coastal areas. These movements are seasonal, and depend both on the availability of resources and on other socio-cultural factors, and the mobility patterns and the distribution of the volume of animals involved change over the course of the year (Apolloni et al., 2019; Bouslikhane, 2015). These movements integrate the region, connect contrasted agro-ecological areas, and, in addition, generate income for many supply chain actors, including producers, traders, transporters, and vendors, and contribute to the food and nutrition security of the region (Valerio, 2020).

As it is, mobility in West Africa is a complex phenomenon involving different temporal scales (from a few days to several months) and spatial scales (from a few kilometers to reach local markets to international transhumance and/or international trade), and whose determinants range from environmental factors, e.g. the availability of natural resources, commercial factors, e.g. market demand and prices, to social factors, such as religious festivals (Apolloni et al., 2019).

In Senegal, livestock production is one of the main economic activities, it involves 28% of the population (ANSD, 2013) and provides almost 4% (ANSD, 2020) of gross national domestic product. Due to the different agro-ecological zones, several production systems co-exist. Senegal is located on the Atlantic Coast axes of the transhumance routes (from Mauritania and Mali to Guinea and Guinea-Bissau) and is involved in international trade movements. Within Senegal, trade follows a strict market hierarchy: from village markets to consumption markets in coastal areas. National transhumance involves movements from the central and southern predominantly agricultural area towards the area of Ferlo in the north.

Like in other West African countries, movements within and towards Senegal vary over the course of the year. This is particularly true of the Tabaski religious festival, an important Muslim festival characterized by the sacrifice of rams, and the Grand Magal of Touba, during which the consumption of beef increases significantly. The two festivals mean imports of livestock increase enormously in a short period of time (Apolloni et al., 2019; Cesaro et al., 2010).

Animal movements also mean pathogens can be introduced and spread at national and international scales. Such pathogens spread very rapidly across national borders and have serious socio-economic and public health consequences. Some of these, such as Contagious Bovine Pleuropneumonia (CBPP), Foot-and-Mouth Disease (FMD), Peste des Petits Ruminants (PPR), and Rift Valley Fever (RVF) are currently a major problem in West Africa (Apolloni et al., 2019; Bouslikhane, 2015; Chaters et al., 2019; Di Nardo et al., 2011).

The porosity of the border, the absence of an animal identification system, together with the lack of coordinated control and surveillance systems hinders the development of a regional surveillance system and increases the risk of epidemics (Apolloni et al., 2019). Understanding mobility patterns, as well as their variations, is of the uttermost importance to optimize surveillance and control systems. Senegal is one of the few countries in West Africa already equipped with a system for mapping and controlling animal movements within its borders. Movements are regulated through the use of sanitary certificates (LPS – *laissez-passer sanitaire*), issued by the veterinary services to livestock transporters every time they move animals. The certificates are also routinely collected and centralized by the veterinary services. The information that can be retrieved from these data provides a snapshot of the livestock mobility network at each period of the year and could be used to develop tools to improve the surveillance system, adapted to the period concerned.

Network-based approaches are widely used in veterinary epidemiology to study the role of animal mobility in the spread of diseases, with the aim of developing effective strategies for disease surveillance and control (Dubé et al., 2009; Motta et al., 2017). Network-based approaches make it possible to depict livestock movements as a spatial network in which the nodes represent villages, administrative units, markets or herds, and a link is created each time at least one animal is moved from one node to another. However, while network methods have been extensively applied to engineer surveillance system in European countries thanks to the existence of vast live animal movement traceability datasets (i.e. Lentz et al. (Lentz et al., 2016) and Schirdewahn et al. (Schirdewahn et al., 2021)), little has been done in West Africa due to the scarcity of such information (Muwonge et al., 2021). Only a few articles that report network analysis in West Africa have been published recently, including Apolloni et al. (Apolloni et al., 2018), and Nicolas et al. (Nicolas et al., 2018) for Mauritania, Jahel et al. (Jahel et al., 2020) for Senegal and Mauritania, and Valerio et al. (Valerio et al., 2020) for the whole West Africa region.

Static network approaches may not be the best way to create effective surveillance and control tools against the spread of infectious diseases, as a static approach can overestimate or underestimate the rate and extent of outbreaks (Masuda & Holme, 2013). The influence of temporality on the structure of the network can significantly affect the spread of a disease, which consequently can only be accurately predicted if the chronology of links is accurately represented (Masuda & Holme, 2013; Williams & Musolesi, 2016).

In this work, we used a temporal network approach to assess the influence of change on the diffusion of animal diseases over time. We used data collected in 2020 by the Senegalese Veterinary Services to build a representation of the network, and adapted tools from complex networks to assess the risk of being infected and the role of different Senegalese areas in spreading infections over the course of the year. To this end, we relied on measures of the "vulnerability" and "reachability" of nodes. Vulnerability gives an indication of the likelihood a node will be infected, while reachability gives an indication of the time to infection. This approach takes changes in the network over time into account as well as the network structure, and differs markedly from the static, individual-centric approaches used in previous risk assessments. The objective of the present work is to provide a theoretical basis for improving the Senegalese surveillance system by identifying different potential geographical spots that contribute to the spread of pathogens at different times of the year and that could be used as sentinel nodes.

## 2. Materials and methods

### 2.1 Study area

Bordering Mauritania to the north, Mali to the east, Guinea and Guinea-Bissau to the south, and the Gambia and the Atlantic Ocean to the west, Senegal occupies an area of 196,722 km2 and in 2020, had an estimated population of more than 16.7 million (World Bank, 2022).

The administration of the Senegalese territory is organized in 14 regions, 45 departments, and 123 *arrondissements* (Ministère de l'Intérieur du Sénégal, 2017). There is a clear contrast between the empty area in the east (hosting around 10% of the human population of Senegal) and the populated and urbanized central and areas in the west, where 90% of human population is concentrated, of which 25% in the Dakar area (ANSD, 2020; World Bank, 2022).

Senegal's climate is very varied and distinct climatic zones are characterized by different levels of rainfall and different types of vegetation. This diverse climate strongly influences the livestock farming sector, whose different farming systems depending on agro-climatic gradients, among other factors (Cesaro et al., 2010).

As mentioned above, the livestock trade is organized in a strict hierarchical system starting at village weekly markets (*Lumo*), the animals are collected by traders to be sold at collection markets before being sent on to consumer markets, where they are sold to be slaughtered. Because there are practically no meat storage facilities, most trade involves live animals. Livestock trade routes converge on the Dakar region, the main consumer market, with stops in smaller markets such as Saint-Louis, Touba, Thiès, and Kaolack. Before reaching the urban markets, the vast majority of the animals originating from northern Senegal, Mauritania, and Mali, are grouped in Dahra, called the "livestock capital" of Senegal. Another collection market in the southeastern part of the country also plays a major role in the livestock trade: Tambacounda, the point of convergence for animals from eastern and southern Senegal, as well as from southern Mauritania and Mali. In addition to the movement of animals for sale, transhumance is widespread in Senegal, both at national scale from the central area to the north (in particular the Ferlo region), and, due to its location, international, from Mali and Mauritania to the Senegalese coast (Apolloni et al., 2019; Cesaro et al., 2010) (SI Figure 1).

## 2.2 Data

In Senegal, a certification system based on a sanitary pass named "*Laissez-Passer Sanitaire*" (LPS) is used to track animal mobility and to map the most important axes of movement in the region. Veterinary posts belonging to the Senegalese Ministry of Livestock and Animal Production provide an LPS each time a herd is moved, the document states the origin of the movement (village, department, region, country), the destination (village, department, region, country), the date, the species and number of animals involved, and the means of transport. Copies of the LPS are centralized and stored in electronic form.

We focused our analyses on movements of cattle and small ruminants (goats and sheep), either separately or together. For analytical purposes, the two were aggregated on the spatial scale of an administrative department (all 45 Senegalese departments are involved in this trade) and on a time scale of one month or one week, depending on the type of analysis: we chose a month as the temporal unit for the general description of the data and for cluster analysis, and a week to simulate the disease spread, as a week is a more realistic unit to study disease propagation.

## 3. Methods
### 3.1 Descriptive and network analyses

We analyzed mobility data using a complex network approach. LPS data were used to build three oriented and weighted networks, one for each species plus one species-independent network: the nodes corresponded to the departments of origin and destination; a direct link existed between two nodes if at least one animal was moved from the department of origin to the destination department; the link was weighted according to the number of animals moved along it.

A cluster analysis was performed to explore the behavior of the different nodes over the study period. The nodes were ranked based on their activity, defined as the effective number of animals traded each month; the number being positive if the inflow of animals was greater than the outflow (*importing behavior*), otherwise negative (*exporting behavior*).

Clustering was performed using HCPC (Hierarchical Clustering on Principal Components), which successively applies three standard methods used in multivariate analyses: (i) Principal Component Analysis (PCA), which identifies the principal components, (ii) hierarchical clustering, which defines the optimal number of clusters of nodes according to their score on the principal components, and (iii) non-hierarchical clustering (in particular the k-means algorithm), which associates a cluster with each node (Celebi, 2015).

To study the structure of the livestock network, we conducted a spatio-temporal analysis of link's frequency, defined as the number of months in the year in which movements occur on the link. We considered a link to be active when at least one trade movement was recorded in a given month. We then categorized the links according to the number of months in which they were active. In particular, we identified four frequency categories, which were, starting from the least frequent: occasional (activity only occurred in one month per year), intermediate (activity occurred in two or three months per year), frequent (activity in four to nine months per year), and backbone (activity in 10 to 12 months).

To compare the risk of diffusion over the course of the year, we used the epidemic threshold $q$ (Volkova et al., 2010). This measure provides information on the minimum probability for a virus to spread throughout the network: the lower the value of the epidemic threshold, the higher the risk of propagation.

For a weighted network, this parameter can be estimated as follows:

$$q_w = \frac{\langle w^{out} \rangle}{\langle w^{in} \times w^{out} \rangle}$$

where ⟨ ⟩ indicates the average value, $w^{in}$ and $w^{out}$ indicate the nodes' in-weight and out-weight, respectively (Nicolas et al., 2018).

Following the procedures of Lancelot et al. (Lancelot et al., 2017) and Nicolas et al. (Nicolas et al. 2018), for each of the three mobility networks considered (All species, Cattle, Small ruminants separately), the epidemic threshold was estimated for each monthly snapshot of the network, to assess the risk of an epidemic occurring over the course of the year.

### 3.2 Simulation of disease spread

Temporality, i.e. the variation in time of the mobility network, affects disease spread. Figure 1 – A shows an example of a temporal network and its static counterpart. The network is composed of seven nodes and eight possible links, whose direction is indicated by the arrows. In this case, the temporal network is characterized by three temporal snapshots that contain the same nodes but different links. A link that is present and active in a snapshot is not necessarily the same in the previous or the following snapshots. If we disregard the information on timing, we obtain an aggregated/static network composed of the same nodes and links as the temporal network, all present and active at the same time. If we simulate an outbreak in the two networks (temporal and static) (Figure 1 – B), we can see that the potential diffusion of the pathogen differs in the two situations. In this case, there is significantly more propagation in the static network than in the temporal one. This happens because, in the temporal network, the disease can only propagate through temporal paths. In other words, if a link connecting an infected node to a susceptible one is active in a specific temporal snapshot, the disease can spread to the latter; conversely, if the link is not active in the temporal window concerned, disease propagation stops.

To study the influence of temporality on disease propagation, we simulated the spread of an animal disease transmitted by direct contact through the livestock mobility networks. We used a SI (Susceptible-Infected) model: the disease was transmitted from an infected node to a susceptible one with a probability of 1, and the infected nodes remained infected for the entire period of analysis, and were consequently able to continue to spread the disease even weeks after being infected. The aim of this procedure was to estimate the number of potentially infected nodes when the underlying structure varied. Because of our focus on the control of transboundary animal diseases, the departments of Mali and Mauritania, which export livestock directly to Senegal, were chosen as sources of the disease, as the majority of Senegalese imports of small ruminants and cattle are from these two countries (Apolloni et al., 2019; Cesaro et al., 2010).

To explore the effect of temporality on the structure of the network, and hence on diffusion of the disease, we compared results obtained with a static representation (in which the structure of the network remains unchanged throughout the year) with results obtained with a temporal representation. In the first case, all the links recorded in the dataset were present at the same time, the time of activation was not taken into account, while in the second case, we included changes in the structure in every week of the study period. To this end, we used temporal path formalism, according to which a temporal path is a sequence of links connecting two nodes with each link in the path coming temporally after the one before it (Masuda & Holme, 2013). This approach enabled us to estimate the infection time: that is the minimum number of timesteps (i.e. weeks) needed to create a temporal path between an infected node and the node under observation.

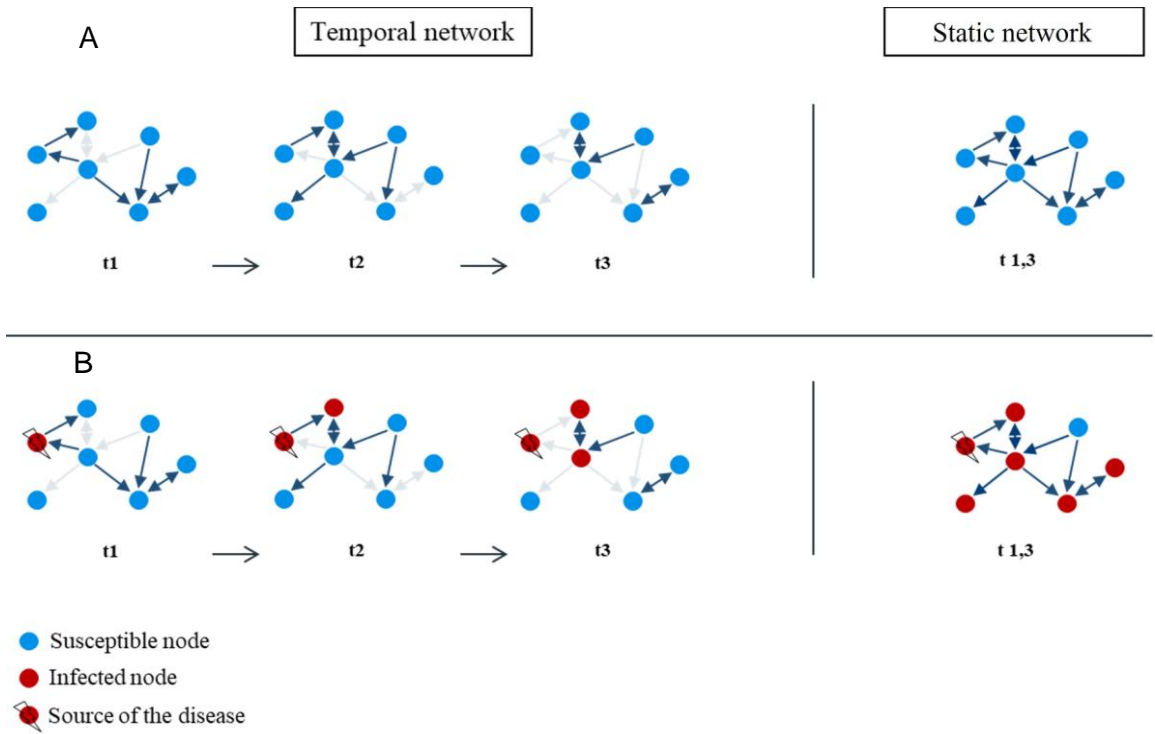

*Figure 1: (A) An example of a directed temporal network and its static counterpart. The dark links are those in the temporal snapshot, while the pale links are those that are possible but are not present in the temporal snapshot. (B) Simple simulation of disease spread in the temporal network (on the left) and the static network (on the right).*

Among all the possible temporal paths between the sources and the other nodes, we decided to consider the "earliest arriving" paths, which represent the first time a node is infected by the disease (Bender-deMoll et al., 2021; Berlingerio et al., 2013). The speed/rate at which a node became infected was estimated by the infection time, i.e. the number of weeks that elapsed between the onset of the disease and the time at which the department concerned was reached for the first time. For static networks, the speed/rate of infection was estimated from the length of the shortest paths, converting the links into temporal units, specifically, weeks. If a node was reached by more than one source, the shortest infection time (for temporal networks) or the shortest path (for static networks) was chosen.

All descriptive analyses and static/temporal network analyses were carried out using R software with the following packages: ggplot2 for graphs (Wickham, 2016), ggplot2 and tmap for maps (Tennekes, 2018); FactoMineR (Lê et al., 2008) and factoextra (Kassambara & Mundt, 2020) for cluster analysis, sna (Butts, 2020), and tsna (Bender-deMoll et al., 2021) for static and temporal network analysis, respectively.

# 4. Results

## 4.1 Summary statistics

The database contained information on 8,861 livestock trade movements from January to December 2020. The network is composed of a total of 88 nodes, corresponding to an Administrative Unit of level 2, of which 45 are Senegalese (Departments), and 590 unique links, i.e. origin-destination combinations. The movements concerned Senegal as the origin and/or destination of 87% of the movements, and eight other countries: Mali (9%), Gambia (2%), Mauritania (1%), Guinea, Guinea-Bissau, Burkina Faso, Niger and Nigeria (<1% each) as either the origin or as the destination of movements. Focusing on Senegal, a total of 6,511 national movements and 2,350 international movements, over respectively 458 and 132 unique links, involving 87,017 cattle and 553,718 small ruminants were recorded in the dataset. Despite the large number of national trades, the majority of animals were moved for the purpose of international trade. More than 95% of these movements were in trucks, which is the most widely used means of transport for animals in the region concerned. More than 600,000 animals were transported by truck, the remainder mainly on foot (Table 1).

The livestock network was analyzed as static but also took temporality into account, which influences the presence/absence of links.

*Table 1: summary of the characteristics of the data analyzed in the study. The number of movements, the number of animals and the number of unique links are given for each species, type of trade, and means of transport.*

|   |   | Trade movements | Headcount | Number of unique links |
|---|---|---|---|---|
| Species |   |   |   |   |
|   | Cattle | 3,186 | 87,017 | 328 |
|   | Small Ruminants | 5,675 | 553,718 | 502 |
| Type |   |   |   |   |
|   | International | 2,350 | 365,903 | 132 |
|   | National | 6,511 | 274,832 | 458 |
| Means of transport |   |   |   |   |
|   | Train | 4 | 170 | 2 |
|   | Truck | 8,239 | 608,816 | 552 |
|   | On foot | 587 | 30,354 | 85 |
|   | Boat | 31 | 1,395 | 9 |

As shown in Figure 2, all Senegalese administrative departments are involved in animal trade either as the origin, the destination, or both. Movements are both national and international, and, while Senegal is the final destination of almost all the trade, many animals are moved not only from other Senegalese departments, but also from Mali and Mauritania, the main exporters, with some departments, particularly in Mali, exporting a significant number of animals (Figure 2– A).

The departments in north-eastern Senegal (Podor, Matam, Kanel, and Ranérou Ferlo), are notable for their high level of animal "exports". Other Senegalese departments (Tambacounda in the south, Koungheul and Gossas in the center, Louga and Kébémer in the north) also export considerable numbers of animals.

Concerning imports, the departments that import the most animals are located in the Dakar region, in particular Pikine, Rufisque, Thiès, and M'bour, where the majority of consumer markets are located. Other Senegalese departments that import large numbers of animals are Saint-Louis in the north, Mbacké and Guinguinéo in the center, Ziguinchor in the south-west, Tambacounda and Sayara in the south-east, on the border with Mali (Figure 2– B).

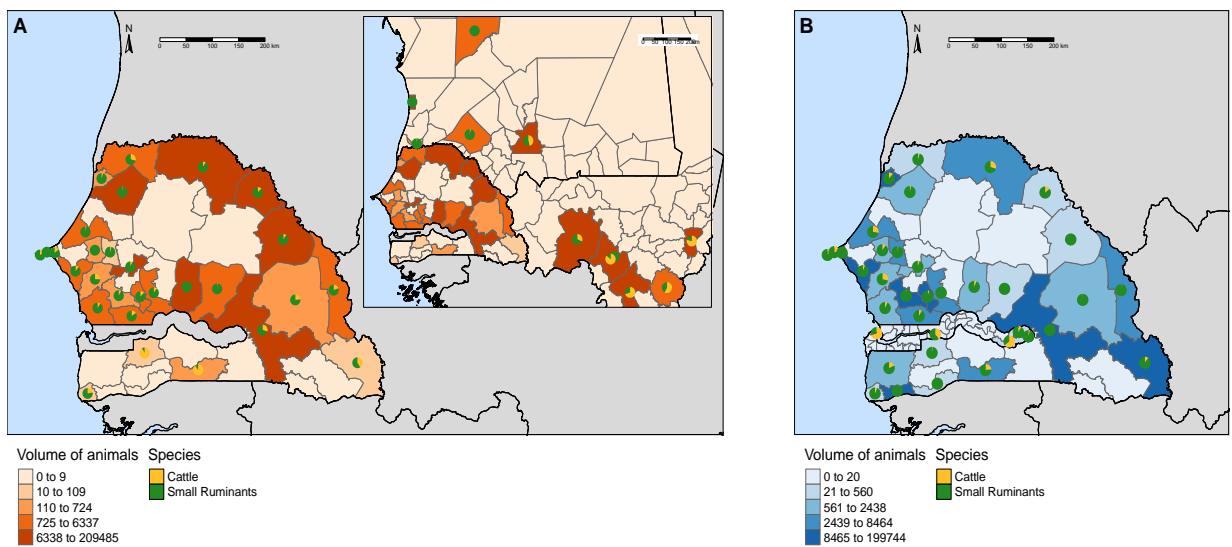

*Figure 2: Distribution of the volume of animals in the administrative departments of Senegal, according to whether the department is the origin (A) or the destination (B) of livestock movement. The miniature pie charts show the percentage of cattle (yellow) and small ruminants (green) in the total number of animals. Quartiles were chosen for the colors representing the volume of animals traded. Only countries that account for at least 1% of exports (A) or imports (B) are shown.*

Figure 3 shows the number of movements and the volume of animals traded in each species (cattle or small ruminants) per month. Overall, movements of animals for the purpose of trade were less frequent in the first six months of the year, but increased in July, particularly trade in small ruminants. Similarly, July was the month with the most trade in small ruminants in the study period, involving more than 300,000 animals. In August and September, the volume of small ruminants decreased, while both the movement and volume of cattle traded increased, overtaking those of small ruminants. In October, November and December, the number of cattle trades decreased, but remained higher than in the rest of the year, while the number and volume of trade in small ruminants increased, although less sharply. In 2020, two important Muslin festivals took place at the end of July (Tabaski) and at the beginning of October (Grand Magal of Touba) and are represented on the chart by a dashed line and a dotted line, respectively (Figure 3).

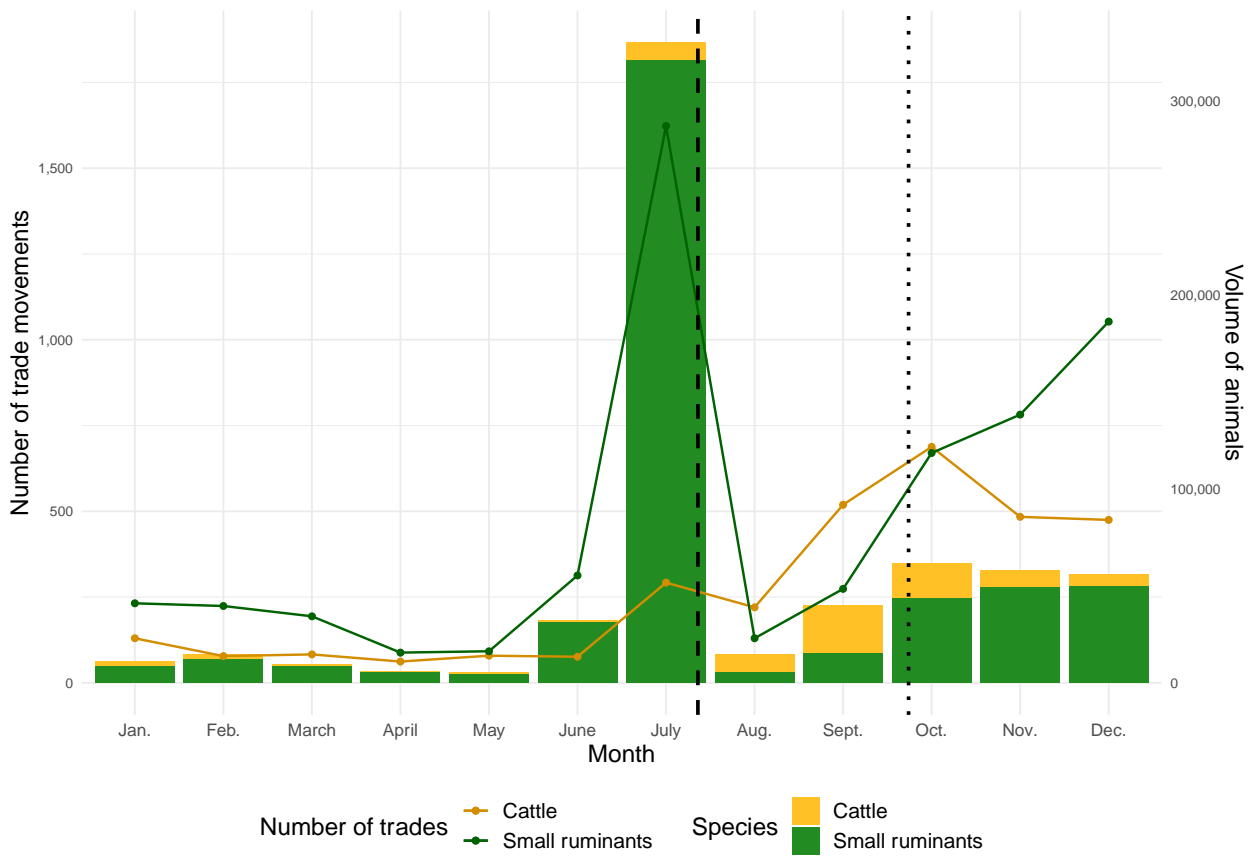

*Figure 3: Number of trade movements (line plot) and volume of livestock traded (bar plot) recorded in 2020, per species and per month. Data concerning cattle are in yellow, data concerning small ruminants are in green. The dashed line represents the Tabaski festival (July 31), the dotted line represents the Grand Magal of Touba festival (October 6).*

In the whole year, trades of small ruminants were concentrated on 503 links and trades in cattle on 329 links, including 242 links trades of both species. Like for small ruminants, the highest number of unique trade links occurred in July, followed by, in decreasing order, December, November and

October, also the months with the highest number of trade links for cattle. The links used by the two species also increased in the last three months of the year (SI Table 1).

Concerning the means of transport, trucks were used for almost all movements of animals for sale throughout the year. The number of movements peaked in July, and, then after a significant drop, started to increase again in September (SI Table 2).

### 4.2 Cluster analysis

Figure 5 shows the nodes of the livestock network in three (3) clusters:

- Cluster 1, composed of 7 nodes and characterized by a "weak"[1] exporting behavior;
- Cluster 2, composed of 61 nodes and characterized by a "strong"[1] exporting behavior;
- Cluster 3, composed by 20 nodes and characterized by a "strong"[1] importing behavior.

Cluster 1 (in red) aggregates seven nodes, of which four are located on the north-eastern border of Senegal (Matam, Podor, Kanel, and Ranérou Ferlo) with high volumes of animals traded, while the other three (Foundiougne, Kaffrine, Gossas) are located on the southern border of Dakar region (Figure 4 – A). However, in September, Cluster 1 imports are "weak", with a slightly less than 3,000 animals imported (Figure 4 – B).

Cluster 2 (in green) aggregates all the foreign nodes, except Banjul (Gambia), and several nodes across Senegal, accounting for a total of 61 nodes out of 88. Cluster 2 is "strong" in terms of volume of animals exported over the year, despite the fact some nodes import more than export. Some nodes that export large numbers of livestock include Bamako (Mali, 208,462) and Nouakchott (Mauritania, 43,472), while Tambacounda (Senegal, 70,845) is a good example of an importing node (Figure 4 – A). The highest number of exports by this cluster occurred in July, when the number of animals exported was slightly under 200,000. (Figure 4 – B).

Cluster 3 (in blue) aggregates 20 nodes of which the majority is concentrated in the Dakar region but includes some nodes in southern Senegal and one foreign node, Banjul (Gambia). Of the nodes located in southern Senegal, two are on the border with Mali (Saraya and Kédougou), while the other four are located farther west. All the nodes in this cluster were characterized by strong import trade, with most imported animals via Pikine (Senegal, 199,703), but also via Thiès (Senegal, 51,939) and Kaolack (Senegal, 46,432) (Figure 4 – A). Reflecting the movements of livestock for export, this cluster shows a peak of imported animals in July, with a volume of around 250,000 animals, and another, less significant increase from September to October (Figure 4 – B).

---

[1] We introduce the terms *importing and exporting behaviour* to indicate those nodes whose net flow of animals (difference between inflow and outflow) through them is positive and negative respectively. Weak and strong refer to magnitude of the net flow (small or large).

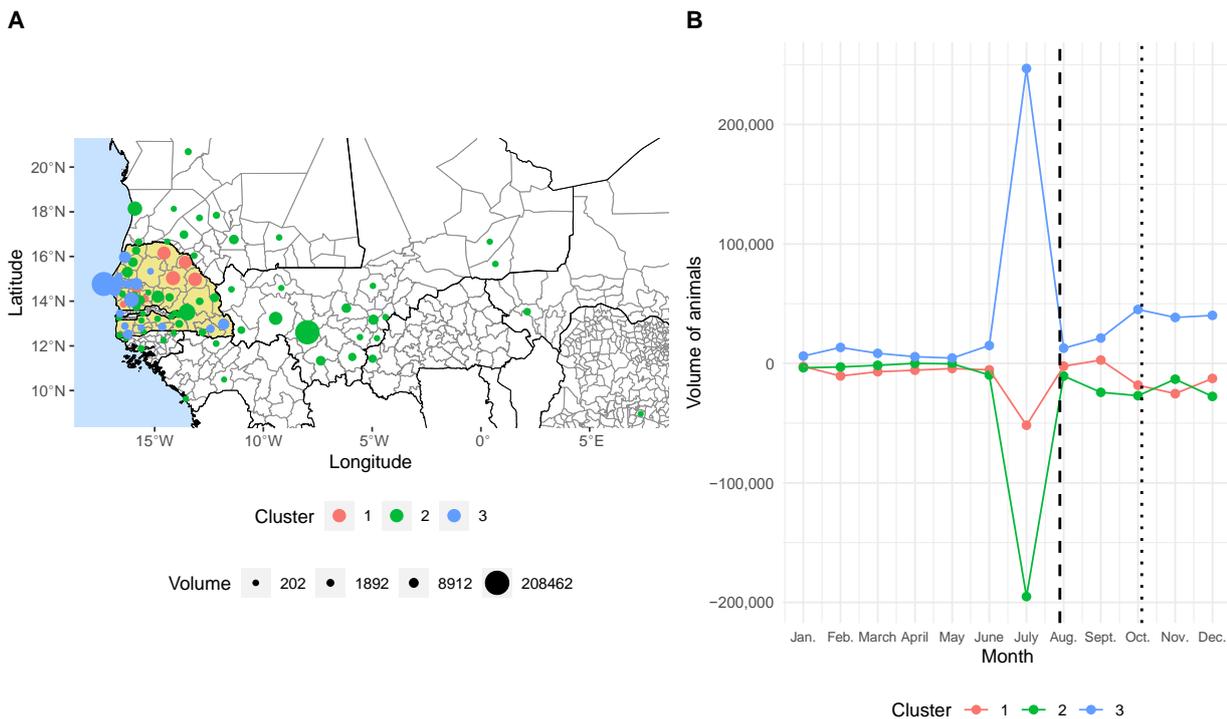

*Figure 4: Clustering of livestock network. (A) Spatial representation of nodes colored according to the cluster to which they belong. The size of each dot indicates the volume of animals traded over the course of the year and the division is made in quantiles. (B) Temporal representation of trade by the three clusters over the course of the year, in terms of the volume of animals traded. Imported animals are represented as positive numbers, exported animals as negative numbers. The dashed line represents the Tabaski festival on July 31, the dotted line represents the Grand Magal of Touba festival on October 6.*

### 4.3 Frequency of links

Figure 5 shows the trade links divided by the frequency of their activity over the course of the year. In general, far more links were characterized by low and very low activity than by very high activity.

The backbone links are three national, short-range connections between north-eastern and north-western nodes: Kanel – Pikine, Ranérou Ferlo – Linguère, Ranérou Ferlo – Mbacké. The majority of frequent links is concentrated in the north of Senegal, where several connections link eastern and western nodes, but some connections link northern and southern nodes. Moreover, some international links are frequent, in particular four originating from Mali and one from Guinea-Bissau. The number of intermediate links is significantly larger than that of the two previous categories, with several connections between Senegal and Mali, and between Senegal and Mauritania. Occasional links are extremely numerous and dense, with several connections in Senegal but also links to all its neighboring countries (Figure 5). The majority of intermediate and occasional links are active in July and October, due to the Tabaski and the Grand Magal of Touba religious festivals. However, considering the whole study period, frequent links are the most common (SI Figure 4).

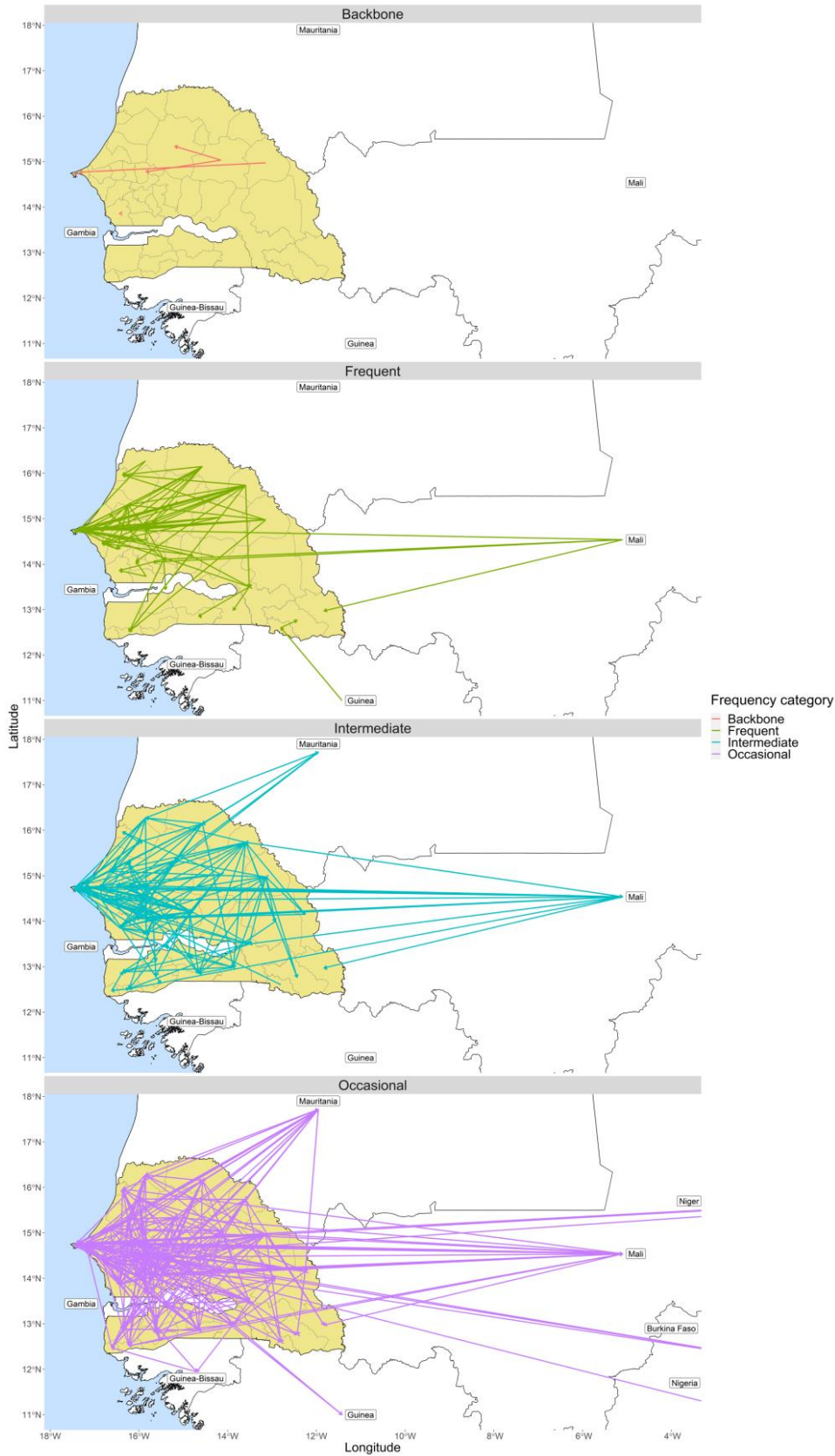

*Figure 5: Geographical representation of the livestock network links, divided by the frequency of their activity over the year. Backbone links were active in more than nine months, frequent links were active between four and nine months, intermediate links were active for two or three months, and occasional links were active for only one month of the study period. We decided to only show the administrative departments of Senegal on these maps. Therefore, concerning national trade, the origin and the destination are both Senegalese departments, while for international trades, they may be a Senegalese department or a central point in a foreign country.*

## 4.4 Epidemic threshold

Overall, the values of the epidemic threshold of all three networks are extremely low, particularly for the combined livestock and the small ruminants network, whose results are almost identical. April is the only month with a significantly higher value than in the rest of the year. On the other hand, the epidemic threshold values of the cattle network (yellow curve in Figure 6), are higher overall, with a value of 1 in April. The lowest values of this network were measured in January, June, and from October until the end of the year (Figure 6).

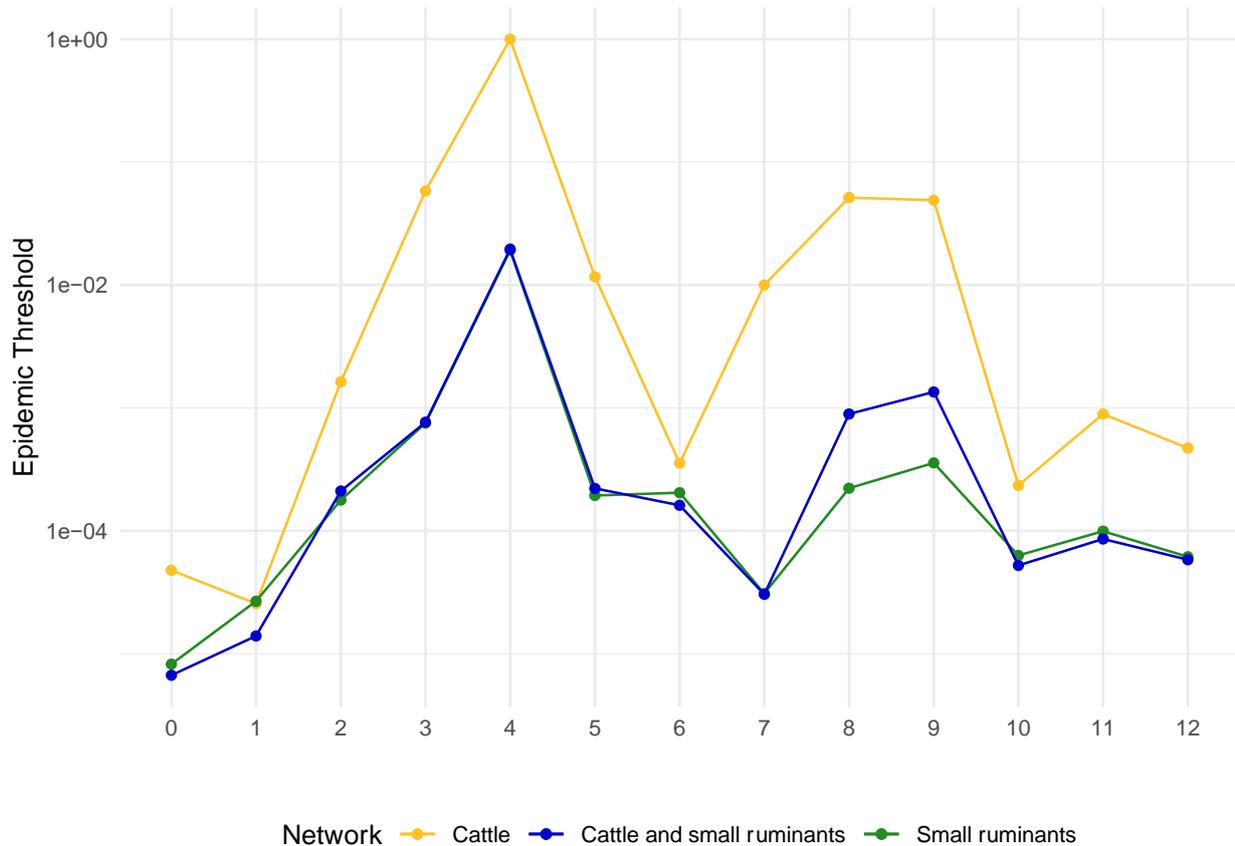

*Figure 6: Logarithmic representation of changes in the epidemic threshold over the course of the year in the three livestock mobility networks. The zero on the left extremity of the horizontal axis identifies the value calculated for the whole year. The small ruminants network is in yellow, the cattle network in green, and the combined livestock network in blue.*

## 4.5 Simulation of disease spread

Maps focused on Senegalese departments were drawn to compare the results of the simulations run on the three networks in an efficient and easily understandable way (Figure 7). To assess the role of changes in the structure of the networks over time, and hence changes in disease spread, we compared the results of a static representation (in the column on the left) with those of a temporal representation

(in the other seven columns). For the static representation, considering the time the outbreak began is meaningless, whereas for the temporal one, it is important, as the network structure can change over time. Therefore, each element in the seven columns representing the temporal networks corresponds to the results of an outbreak that began in a specific week of the year (the number of the week is given in the header of each map).

The departments are colored according to their *infection time*, i.e. the length of the period before they were reached by the virus. For each scenario, the *infection time* was estimated as the time (number of weeks) elapsed since the outbreak of the epidemic. We created four categories of infection time, each category is shown in a different color: red for departments infected less than one month from the beginning of the disease spread (less than 5 weeks), orange for departments infected after 1-2 months (between 5 and 9 weeks), yellow for departments infected after more than two months (more than 9 weeks), and green for those never reached by the disease. If a node was reached by infections from several sources, the shortest *infection time* was chosen.

For the static networks, we considered the number of links in the shortest path between the source and the node as weeks: red for the shortest paths with less than 5 links, orange for paths with between 5 and 9 links, yellow for paths with more than 9 links, and green for nodes that were never reached by the disease.

The results presented are those of the simulation of a disease spreading from Mali, the origin of most animals imported into Senegal in 2020. For the temporal networks, we present only a few weeks characterized by activity, in order to be able to simultaneously show changes over time and differences between the three networks. The complete results of the spread of a disease from Mali plus for a disease spreading from Mauritania can be found in Supplementary Information (SI Figure 5 – 10).

In general, in all three networks, maps representing aggregate networks strongly overestimated both the quantity of potentially infected nodes and the earliness of infection, compared to those of temporal networks. In addition, there is a difference in the potential sanitary risk between the cattle temporal network and small ruminants temporal network, the latter showing on average wider and potentially greater disease propagation. However, the combined network of cattle and small ruminants (the livestock network) is under the greatest sanitary risk.

Figure 7 also shows that, particularly for the livestock network and the small ruminants network, the periods around religious festivals (weeks 30 and 31 for the Tabaski, and weeks 40 and 41 for the Grand Magal of Touba) are characterized by a large number of infected departments, some of which are infected early.

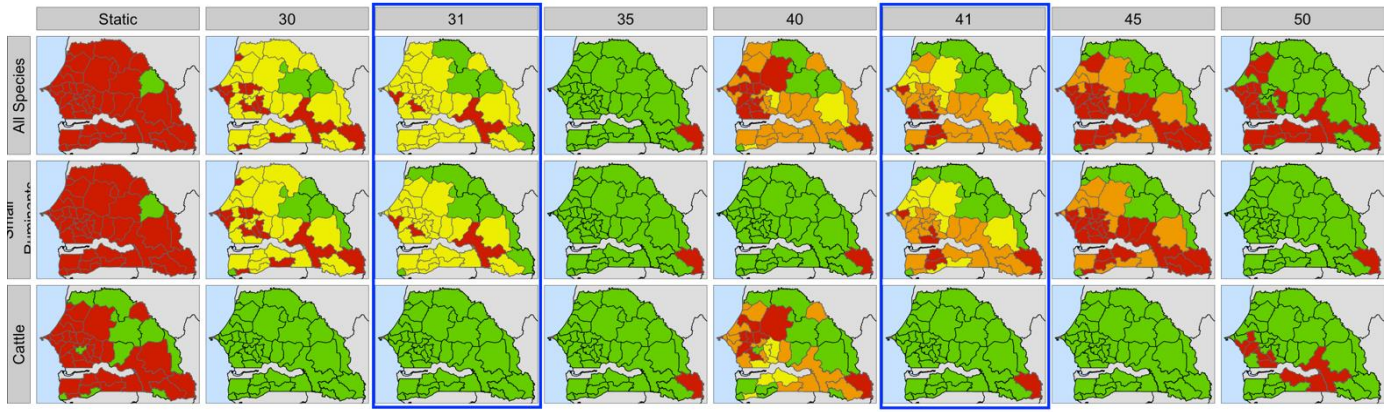

*Figure 7: Geographical representation of infection time in the case of disease propagation from Mali. For each mobility network, the first column on the left represents the spread in the static network, the other seven columns represent the seven worst scenarios of transmission if that specific week represents the beginning of the disease spread. The colors indicate the infection time of the disease: red for less than one month, orange for less than two months, yellow for more than two months, green for nodes that have never been touched in one year time. For the static networks in the first column on the left, the colors are based on the number of links in the path: up to 5 in red, green for nodes that have never been touched in one year time. The squares outlined in blue identify the week of the Tabaski festival (week 31) and the week of the Grand Magal of Touba festival (week 41).*

## 5. Discussion and conclusions

In 2020, during the COVID-19 pandemic, several restrictive measures were introduced in Senegal that affected both human and livestock mobility. Borders were closed for both humans and animals in March 2020 and, at the same time, movements between regions were regulated. To supply markets and families in preparation for the Tabaski festival on July 31, borders were reopened 45 days before Tabaski and measures were eased for national and international movement (lettre circulaire n° 01806 PR/MESG/CT-PSS du 17 juin 2020). Similar decisions were taken on the occasion of the Grand Magal of Touba, a religious pilgrimage during which a large number of cattle in particular are sold and consumed. The application of these restrictions had a huge effect on the structure of the networks and on the risk of introduction and diffusion of pathogens, as did re-opening the border. To assess the impact of these measures on the spread of livestock diseases, and for possible future use, we used tools from temporal network theory to identify the area with the highest risk of introduction. It is important to note that normally, there are other religious festivals, like Gamou of Tivaouane, in addition to the Grand Magal of Touba, but these were cancelled due to COVID-19 pandemic.

In our study, we considered the diffusion of a generic direct animal disease transmission and estimated the vulnerability and the reachability of nodes when the underlined network changes over time. In this way, we were able to identify Senegalese departments that could be infected at the earliest stage of an epidemic. With a few modifications, our approach could be extended to include vector-borne diseases.

The structure of the Senegalese livestock network varies widely over the course of the year due to the seasonality of transhumance and the effect of religious festivals (Apolloni et al., 2019; Jahel et al., 2020) but, in 2020, these effects were exacerbated by the restrictive measures introduced as a result of Covid-19. In fact, around June and July, we noted a pickup in the movement and exchange of animals (mainly small ruminants) mainly due to the easing of the restrictive measures in preparation for the Tabaski festival and (mainly cattle) for the Grand Magal of Touba festival. We also noted that the dynamics of the small ruminants trade strongly drive the dynamics of the network as a whole.

Dakar is the main consumption area of Senegal because almost a quarter of the population of the country live in the city. Consequently, the main markets of Dakar and Pikine (at the entrance of Dakar) are the main destination of livestock movements. In particular, regular movements occur between the areas of Kanel, Ranerou Ferlo, Dahra and the Senegalese capital. In these areas, there is a high concentration of collection markets (local name *luma*), where traders frequently buy animals to be sold directly to Dakar, or to the other collection markets in Dahra or Thiès before being sent on to the capital city. Overall, the majority of northern links end in the Dakar region, or in some smaller but nevertheless important markets such as Saint-Louis, Thiès, Mbour, but also Ziguinchor in the south. Some links in the southeast start from Tambacounda, an important point of convergence for animals from eastern Senegal, as well as from Mauritania and Mali. Our analysis revealed that the role played by the different departments changes over the course of the year. Locations that are idle for a large part of the year become active during the Tabaski period and continue to be active until the end of the year, in particular, departments that produce small ruminants. Occasional and intermediate links that are active a few times a year, are usually located near festival centers, to support the increased supply of livestock, thereby increasing the sanitary risk.

Analysis of the threshold parameters showed that the network is prone to disease spread, but that the risk fluctuates over the course of the year. The risk increases significantly on the occasion of festivals due to the introduction of large numbers of animals and the creation of new commercial routes, and diseases can then spread easily across the network. However, the potential infected areas, and the reachable time does not remain stable over the course of the year and this information cannot be captured using a static representation of the network. In fact, a static representation of the mobility patterns may largely overestimate the speed and the extent of disease diffusion: when the simple static approach is used, diseases appear to spread throughout the country in less than a month, whereas temporal analysis shows that reachability and vulnerability of departments varies over the course of the year. In most cases, and depending on the species involved, few departments are reached in a month, although during the months around Tabaski and Grand Magal, the number of departments that can be reached increases drastically, and for some departments (like Dakar, Thiès, Tambacounda and Dahra) where the main markets are located, and at the border, this risk is even higher. These results could be of great interest not only for risk-based surveillance but also for optimizing the distribution of resources and personnel needed for control at specific times of the year by focusing on the areas that are most likely to be reached. The fact that departments located at the border are most prone to early infection, means that sanitary control at the border should be strengthened and surveillance and control measures should be harmonized at regional level.

Previous works already underlined the importance of mobility and of data collection as a tool to improve surveillance and control in Africa (Chaters et al., 2019; Motta et al., 2017; Nicolas et al., 2018). Our work fits into this strand, emphasizing the importance of collecting data on animal mobility on a regular basis in order to retrieve information on structural changes. The objective of the present study is to provide theoretical tools to assess the importance of network dynamics when planning control and surveillance policies. A more detailed analysis focused on specific diseases and that accounts for volume distribution may reduce the list of departments to monitor. To this end, further simulations are needed and their results will depend to a large extent on the characteristics of the disease concerned, e.g. it transmissibility and incubation period, that could shape the spatio-temporal pattern of the epidemics and hence the involvement of the different departments. Future works should thus also consider stochastic models like Kim et al. (2018) (Kim et al., 2018) for specific diseases. In the model we used here, we aggregated data at the spatial scale of an administrative department, based on the assumption that the diffusion within a department is homogeneous. In practice, the presence of markets or transhumance corridors could attract movements in specific parts of the department, thereby increasing the risk in certain locations over the risk in other parts of the same department. Data on movements within departments were rare in our dataset (because of the way data were collected) and further field studies are recommended to collect data at a finer scale.

## 6. Acknowledgments

The authors are grateful to Dr Mbargou Lô, head of the veterinary Services Directorate, and Dr Mathioro Fall, head of Animal Health Protection Division. The authors acknowledge the support of Yves Amevoin, Alioune Ka, Fallou Niakh, and Khady Ndiaye, and all those who assisted in the LPS collection. We thank all the donors who supported the CGIAR Livestock research program through their contributions to the CGIAR trust fund.

## 7. Funding

This study was partially funded by the Project Eco-PPR (European Commission through the International Fund for Agricultural Development (grant number 2000002577) and the CGIAR Livestock research program, RVF OIE twinning program (CIRAD-ISRA) granted through the EBO-SURSY project (European Union FOOD/2016/379-660). The funders had no role in study design, data collection and analysis, decision to publish, or preparation of the manuscript.

## 8. Conflict of interest statement

All authors declare that they have no conflicts of interest.

## 9. References

ANSD. (2013). Recensement Général de la Population et de l'Habitat, de l'Agriculture et de l'Elevage (RGPHAE).

ANSD. (2020). Situation économique et sociale du Sénégal Ed. 2017/2018 (p. 413).

Apolloni, A., Corniaux, C., Coste, C., Lancelot, R., & Toure, I. (2019). Livestock Mobility in West Africa and Sahel and Transboundary Animal Diseases. In Transboundary Animal Diseases in Sahelian Africa and Connected Regions (p. 31:52). Springer International Publishing. https://doi.org/10.1007/978-3-030-25385-1

Apolloni, A., Nicolas, G., Coste, C., EL Mamy, A. B., Yahya, B., EL Arbi, A. S., Gueya, M. B., Baba, D., Gilbert, M., & Lancelot, R. (2018). Towards the description of livestock mobility in Sahelian Africa: Some results from a survey in Mauritania. PLoS ONE, 13(1). https://doi.org/10.1371/journal.pone.0191565

Bender-deMoll, S., Morris, M., & Moody, J. (2021, aprile 23). Tools for Temporal Social Network Analysis [R package tsna version 0.3.3]. Comprehensive R Archive Network (CRAN). https://CRAN.R-project.org/package=tsna

Berlingerio, M., Coscia, M., Giannotti, F., Monreale, A., & Pedreschi, D. (2013). Evolving networks: Eras and turning points. Intell. Data Anal. https://doi.org/10.3233/IDA-120566

Bossard, L. (2009). Regional Atlas on West Africa. Éditions OCDE. https://doi.org/10.1787/9789264056763-en

Bouslikhane, M. (2015). CROSS BORDER MOVEMENTS OF ANIMALS AND ANIMAL PRODUCTS AND THEIR RELEVANCE TO THE EPIDEMIOLOGY OF ANIMAL DISEASES IN AFRICA. OIE Regional Commission.

Butts, C. T. (2020, ottobre 6). Tools for Social Network Analysis [R package sna version 2.6]. Comprehensive R Archive Network (CRAN). https://CRAN.R-project.org/package=sna

Celebi, M. E. (A c. Di). (2015). Partitional Clustering Algorithms. Springer International Publishing. https://doi.org/10.1007/978-3-319-09259-1

Cesaro, J. D., Magrin, G., & Ninot, O. (2010). Atlas de l'elevage au Senegal: Commerces et territoires. PRODIG. http://publications.cirad.fr/une_notice.php?dk=558823

Chaters, G., Johnson, P., Cleaveland, S., Crispell, J., De Glanville, W., Doherty, T., Matthews, L., Mohr, S., Nyasebwa, O., Rossi, G., Salvador, L., Swai, E., & Kao, R. (2019). Analysing livestock network data for infectious disease control: An argument for routine data collection in emerging economies. Philosophical Transactions of the Royal Society B: Biological Sciences, 374, 20180264. https://doi.org/10.1098/rstb.2018.0264

Di Nardo, A., Knowles, N. j, & Paton, D. j. (2011). Combining livestock trade patterns with phylogenetics to help understand the spread of foot and mouth disease in sub-Saharan Africa, the Middle East and Southeast Asia. 30(1), 63. https://doi.org/10.20506/rst.30.1.2022

Dubé, C., Ribble, C., Kelton, D., & McNab, B. (2009). A Review of Network Analysis Terminology and its Application to Foot-and-Mouth Disease Modelling and Policy Development. Transboundary and Emerging Diseases, 56(3), 73–85. https://doi.org/10.1111/j.1865-1682.2008.01064.x

Jahel, C., Lenormand, M., Seck, I., Apolloni, A., Toure, I., Faye, C., Sall, B., Lo, M., Diaw, C. S., Lancelot, R., & Coste, C. (2020). Mapping livestock movements in Sahelian Africa. Scientific Reports, 10. https://doi.org/10.1038/s41598-020-65132-8

Kassambara, A., & Mundt, F. (2020, aprile 1). Extract and Visualize the Results of Multivariate Data Analyses [R package factoextra version 1.0.7]. Comprehensive R Archive Network (CRAN). https://CRAN.R-project.org/package=factoextra

Kim, Y., Dommergues, L., M'sa, A. B., Mérot, P., Cardinale, E., Edmunds, J., Pfeiffer, D., Fournié, G., & Métras, R. (2018). Livestock trade network: Potential for disease transmission and implications for risk-based surveillance on the island of Mayotte. Scientific Reports, 8(1), 11550. https://doi.org/10.1038/s41598-018-29999-y

Lancelot, R., Béral, M., Rakotoharinome, V. M., Andriamandimby, S.-F., Héraud, J.-M., Coste, C., Apolloni, A., Squarzoni-Diaw, C., de La Rocque, S., Formenty, P. B. H., Bouyer, J., Wint, G. R. W., & Cardinale, E. (2017). Drivers of Rift Valley fever epidemics in Madagascar. Proceedings of the National Academy of Sciences, 114(5), 938–943. https://doi.org/10.1073/pnas.1607948114

Lê, S., Josse, J., & Husson, F. (2008). FactoMineR: A Package for Multivariate Analysis. Journal of Statistical Software, 25(1), 1–18. https://doi.org/10.18637/jss.v025.i01

Lentz, H. H. K., Koher, A., Hövel, P., Gethmann, J., Sauter-Louis, C., Selhorst, T., & Conraths, F. J. (2016). Disease Spread through Animal Movements: A Static and Temporal Network Analysis of Pig Trade in Germany. PloS One, 11(5), e0155196. https://doi.org/10.1371/journal.pone.0155196

Masuda, N., & Holme, P. (2013). Predicting and controlling infectious disease epidemics using temporal networks. F1000Prime Reports, 5, 6. https://doi.org/10.12703/P5-6

Ministère de l'Intérieur du Sénégal. (2017). Politique de Gouvernance intérieure | Ministère de l'Intérieur. https://interieur.sec.gouv.sn/administration-territoriale/politique-de-gouvernance-interieure

Missohou, A., Nahimana, G., Ayssiwede, S. B., & Sembene, M. (2016). Elevage caprin en Afrique de l'Ouest: Une synthèse. Revue d'élevage et de médecine vétérinaire des pays tropicaux, 69(1), 3. https://doi.org/10.19182/remvt.31167

Motta, P., Porphyre, T., Handel, I., Hamman, S. M., Ngu Ngwa, V., Tanya, V., Morgan, K., Christley, R., & Bronsvoort, B. M. deC. (2017). Implications of the cattle trade network in Cameroon for regional disease prevention and control. Scientific Reports, 7. https://doi.org/10.1038/srep43932

Muwonge, A., Bessell, P. R., Porphyre, T., Motta, P., Rydevik, G., Devailly, G., Egbe, N. F., Kelly, R. F., Handel, I. G., Mazeri, S., & Bronsvoort, B. M. deC. (2021). Inferring livestock movement networks from archived data to support infectious disease control in developing countries. https://doi.org/10.1101/2021.03.18.435930

Nicolas, G., Apolloni, A., Coste, C., Wint, G. R. W., Lancelot, R., & Gilbert, M. (2018). Predictive gravity models of livestock mobility in Mauritania: The effects of supply, demand and cultural factors. PLoS ONE, 13(7). https://doi.org/10.1371/journal.pone.0199547

Schirdewahn, F., Lentz, H. H. K., Colizza, V., Koher, A., Hövel, P., & Vidondo, B. (2021). Early warning of infectious disease outbreaks on cattle-transport networks. PLOS ONE, 16(1), e0244999. https://doi.org/10.1371/journal.pone.0244999

Tennekes, M. (2018). tmap: Thematic Maps in R. Journal of Statistical Software, 84(6). https://doi.org/10.18637/jss.v084.i06

Valerio, V. C. (2020). The structure of livestock trade in West Africa (West African Papers Fasc. 29; West African Papers, Vol. 29). https://doi.org/10.1787/f8c71341-en

Valerio, V. C., Walther, O. J., Eilittä, M., Cissé, B., Muneepeerakul, R., & Kiker, G. A. (2020). Network analysis of regional livestock trade in West Africa. PLoS ONE, 15(5). https://doi.org/10.1371/journal.pone.0232681

Volkova, V. V., Howey, R., Savill, N. J., & Woolhouse, M. E. J. (2010). Sheep Movement Networks and the Transmission of Infectious Diseases. PLoS ONE, 5(6), e11185. https://doi.org/10.1371/journal.pone.0011185

Wickham, H. (2016). ggplot2: Elegant Graphics for Data Analysis (2nd ed. 2016). Springer International Publishing : Imprint: Springer. https://doi.org/10.1007/978-3-319-24277-4

Williams, M. J., & Musolesi, M. (2016). Spatio-temporal networks: Reachability, centrality and robustness. Royal Society Open Science, 3(6), 160196. https://doi.org/10.1098/rsos.160196

World Bank. (2022). Population Senegal [Text/HTML]. World Bank. https://www.worldbank.org/en/country/senegal/overview

# Supplementary information

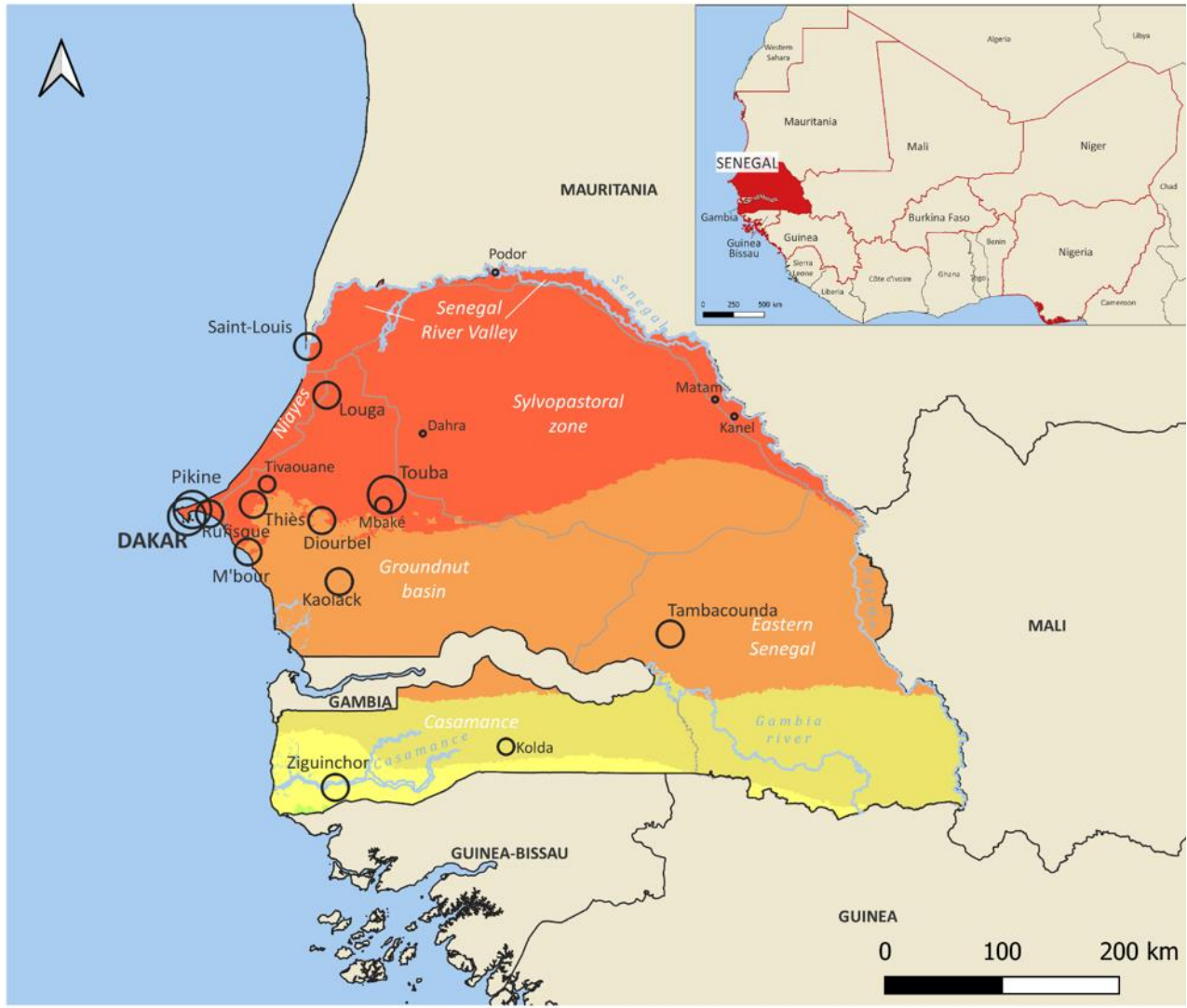

*SI Figure 1: Map of Senegal colored based on the aridity index. The inset map shows all the countries involved in livestock mobility network.*

|  |  | Small Ruminants | Cattle | Links in common |
|---|---|---|---|---|
| Whole year |  | 503 | 329 | 242 |
| Months |  |  |  |  |
|  | January | 42 | 27 | 13 |
|  | February | 52 | 27 | 22 |
|  | March | 59 | 30 | 19 |
|  | April | 26 | 18 | 12 |
|  | May | 27 | 22 | 11 |
|  | June | 71 | 28 | 20 |
|  | July | 202 | 47 | 37 |
|  | August | 29 | 21 | 6 |
|  | September | 24 | 41 | 14 |
|  | October | 131 | 123 | 66 |
|  | November | 163 | 113 | 69 |
|  | December | 184 | 128 | 82 |

*SI Table 1: Number of unique trade links in the small ruminant network and in the cattle network. The values are represented divided by month, while the "whole year" line corresponds to the number of unique links considering the network as static. The last column shows the number of links that are used by the two species.*

|  |  | Truck | Others | | | Links in common |
|---|---|---|---|---|---|---|
|  |  |  | Water | Walking | Train |  |
| Whole year |  | 552 | 9 | 85 | 2 | 55 |
| Months |  |  |  |  |  |  |
|  | January | 49 | 0 | 8 | 0 | 1 |
|  | February | 56 | 0 | 3 | 0 | 2 |
|  | March | 67 | 0 | 3 | 0 | 0 |
|  | April | 31 | 0 | 1 | 0 | 0 |
|  | May | 38 | 0 | 0 | 0 | 0 |
|  | June | 73 | 0 | 9 | 0 | 3 |
|  | July | 207 | 0 | 11 | 0 | 6 |
|  | August | 43 | 0 | 3 | 0 | 2 |
|  | September | 42 | 3 | 11 | 0 | 5 |
|  | October | 178 | 1 | 17 | 1 | 9 |
|  | November | 178 | 3 | 33 | 0 | 7 |
|  | December | 219 | 3 | 20 | 1 | 11 |

*SI Table 2: Number of unique trade links according to the means of transport. The values are given per month, while the "whole year" line corresponds to the number of unique links considering the network as static. The last column shows the number of links that are shared by the movements made by truck and those made by all other types of transport, including on foot.*

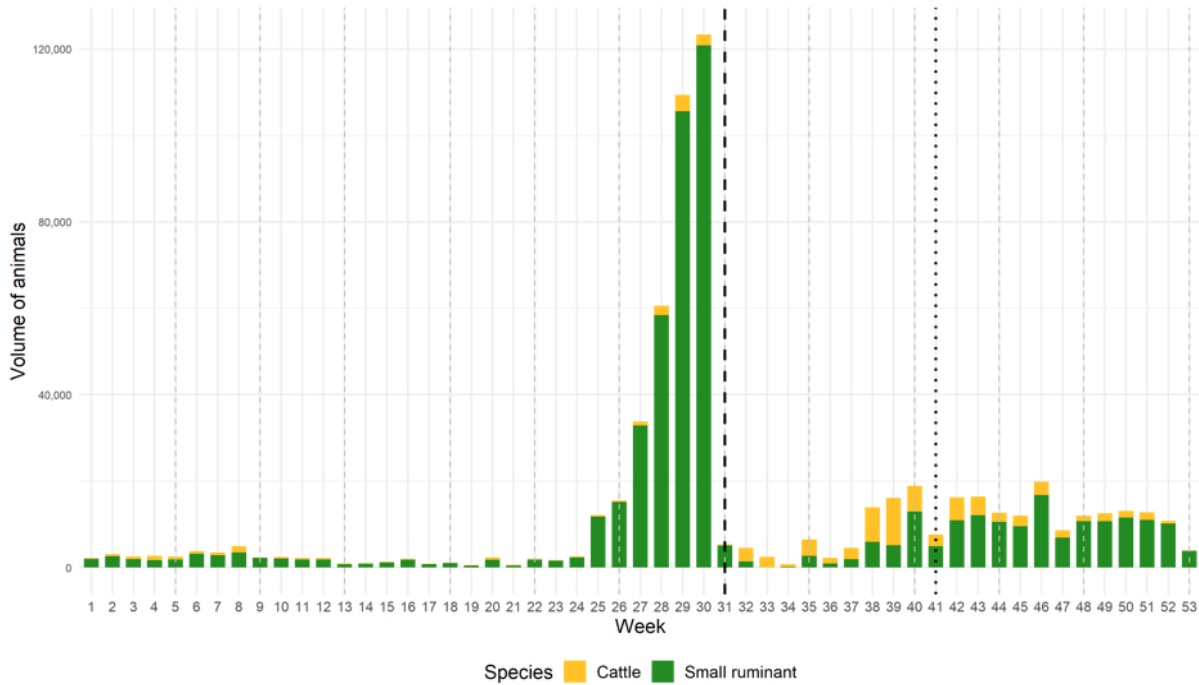

SI Figure 2: volume of livestock traded in 2020 divided by week. Cattle are shown in yellow and small ruminants are in green. The black dashed line represents the day of the Tabaski festival (July 31), the black dotted line represents the day of the Grand Magal of Touba (October 6). The months are indicated by the gray dashed lines.

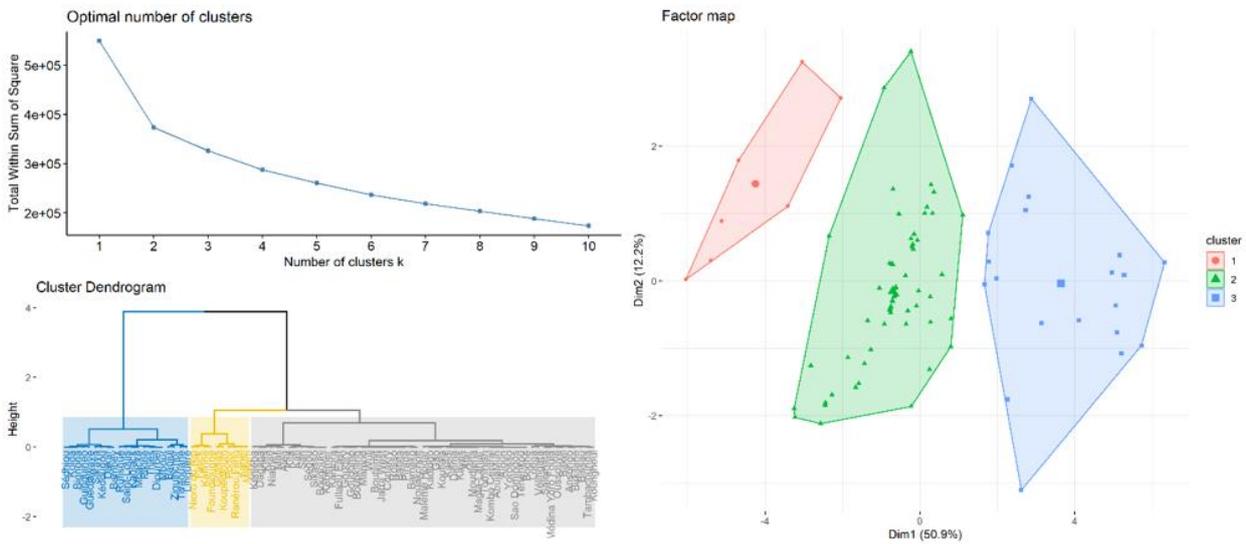

SI Figure 3: Graphical visualization of methods chosen to assess the number of clusters: (A) Elbow method, (B) cluster dendrogram, and (C) cluster division with the HCPC (Hierarchical Clustering on Principal Components) function of the FactoMineR package.

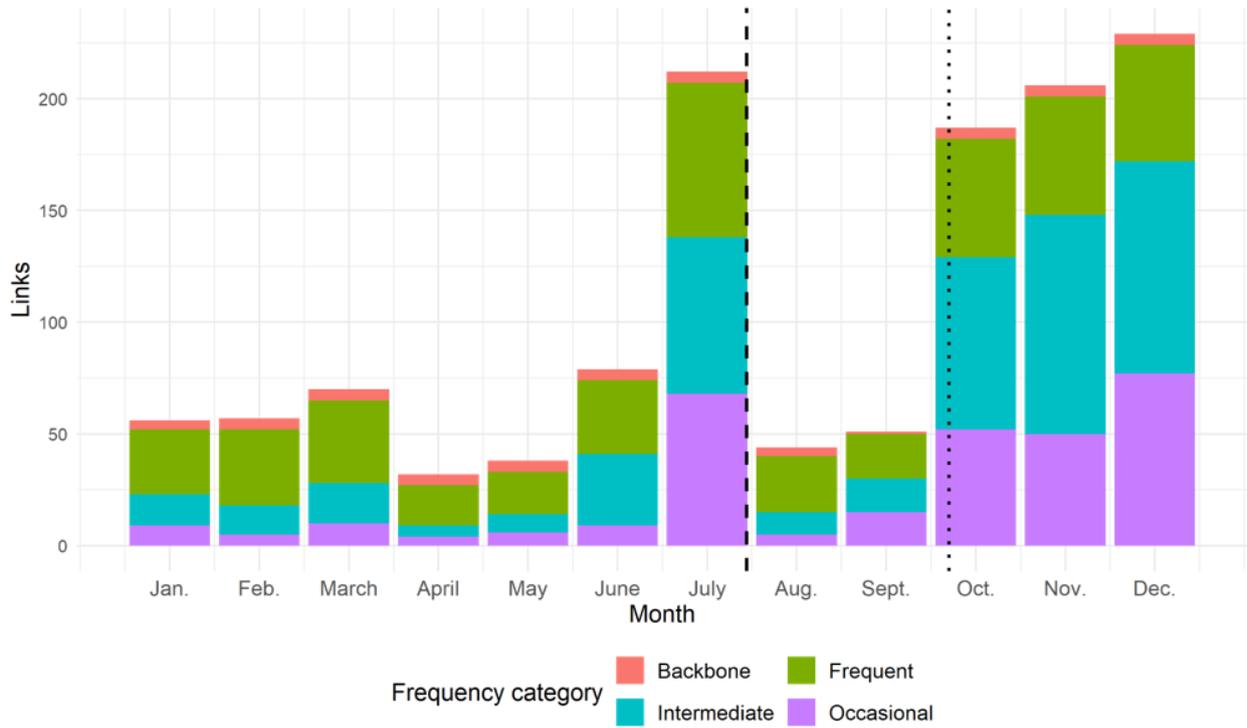

SI Figure 4: Activity of the links over the course of the year. For each month, the quantity of active links is shown, colored according to their frequency. Backbone links are those active more than nine months a year, frequent links are those active from four to nine months, intermediate links are those active two or three months, and occasional links are only active one month. The orange line represents the Tabaski festival (July 31), the violet line represents the Grand Magal of Touba festival (October 6).

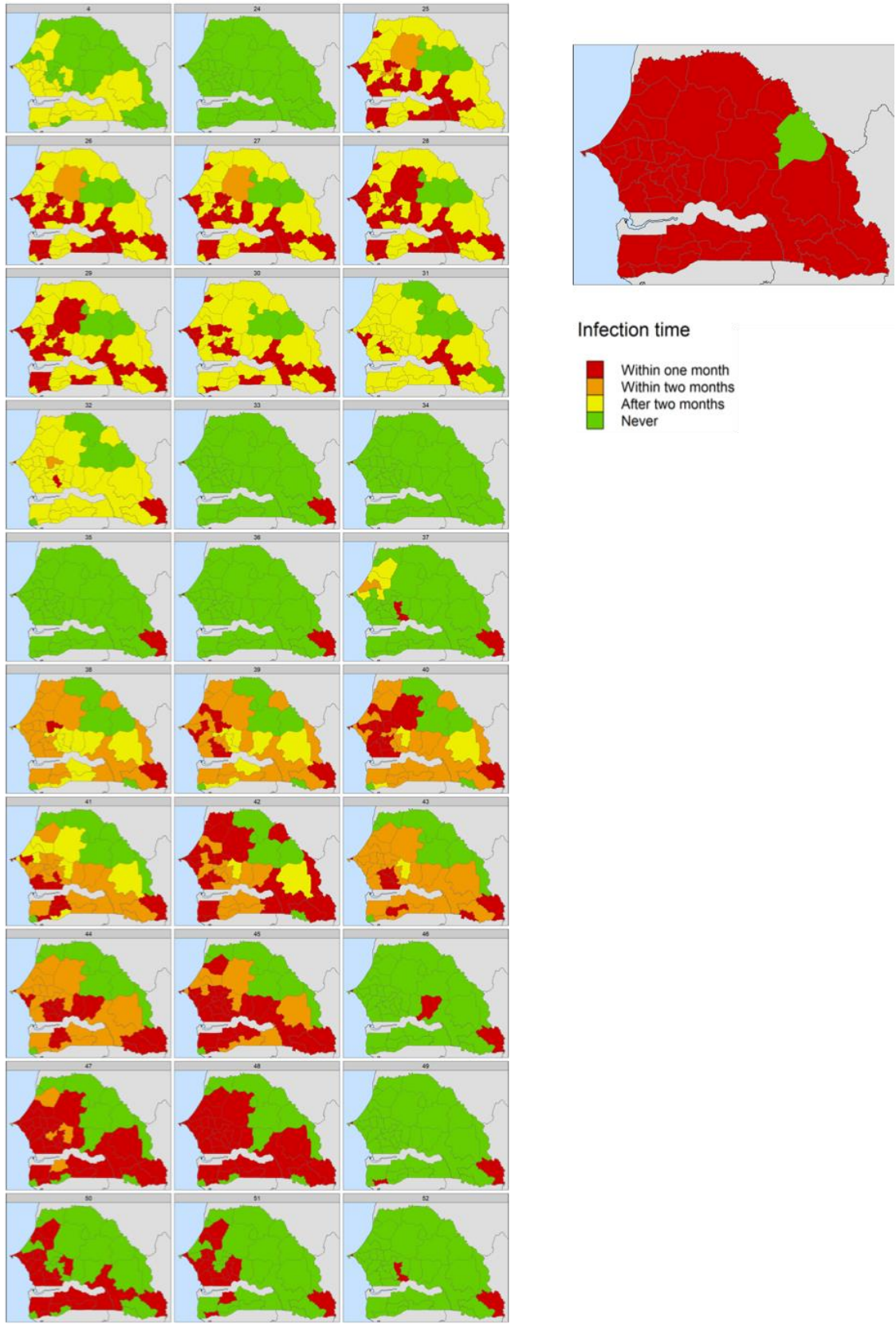

*SI Figure 5: Geographical representation of infection time, in the case of a disease propagated from Mali through the livestock network. Temporal network on the left, static network on the right. For the static network, the colors are based on the links in the path: up to 5 in red, between 5 and 9 in orange, more than 9 in yellow. Nodes that have never been reached are in green.*

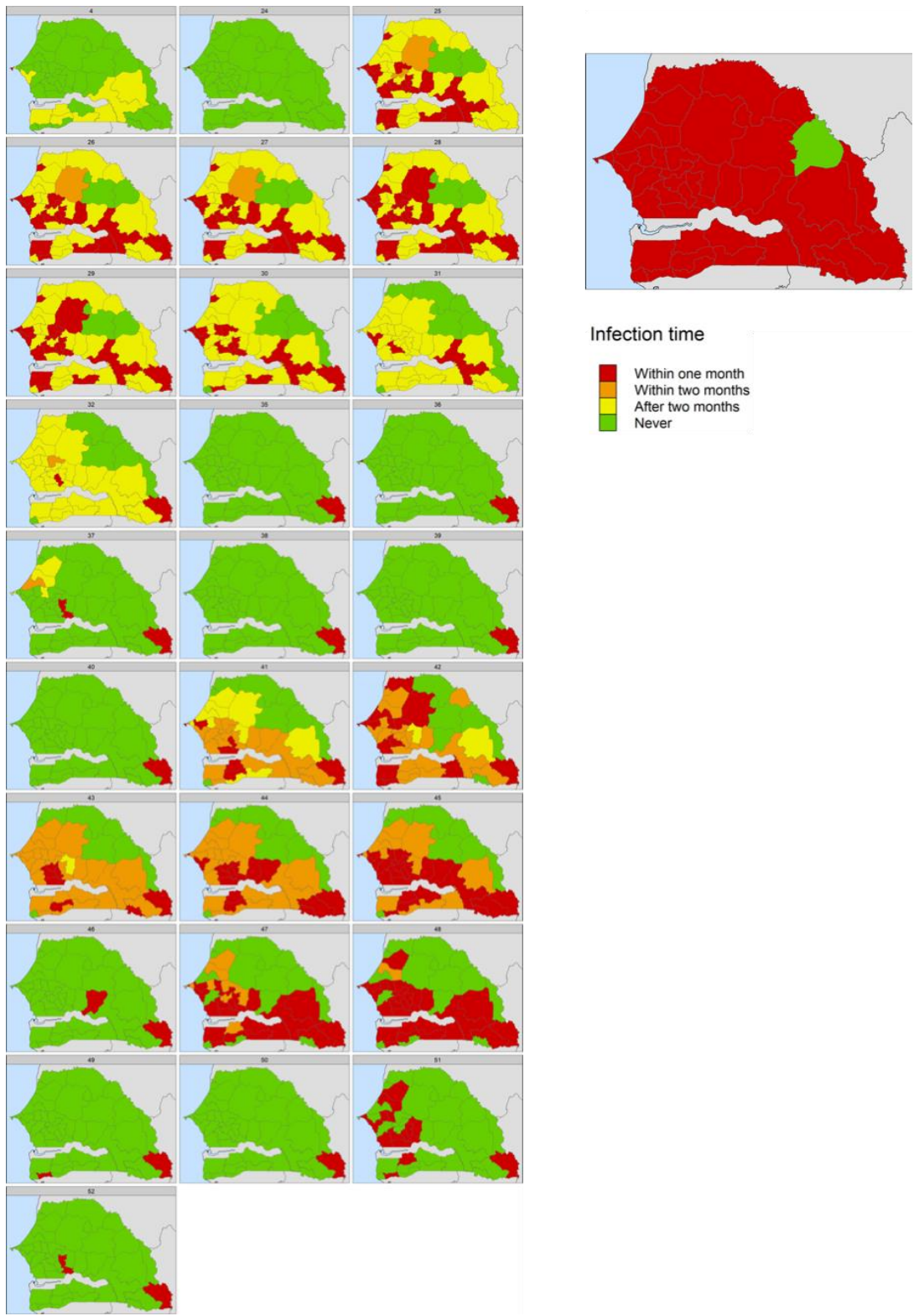

*SI Figure 6: Geographical representation of infection time, in the case of a disease propagated from Mali through the small ruminant network. Temporal network on the left, static network on the right. For the static network, the colors are based on the links in the path: up to 5 in red, between 5 and 9 in orange, more than 9 in yellow. Nodes that have never been touched are colored green.*

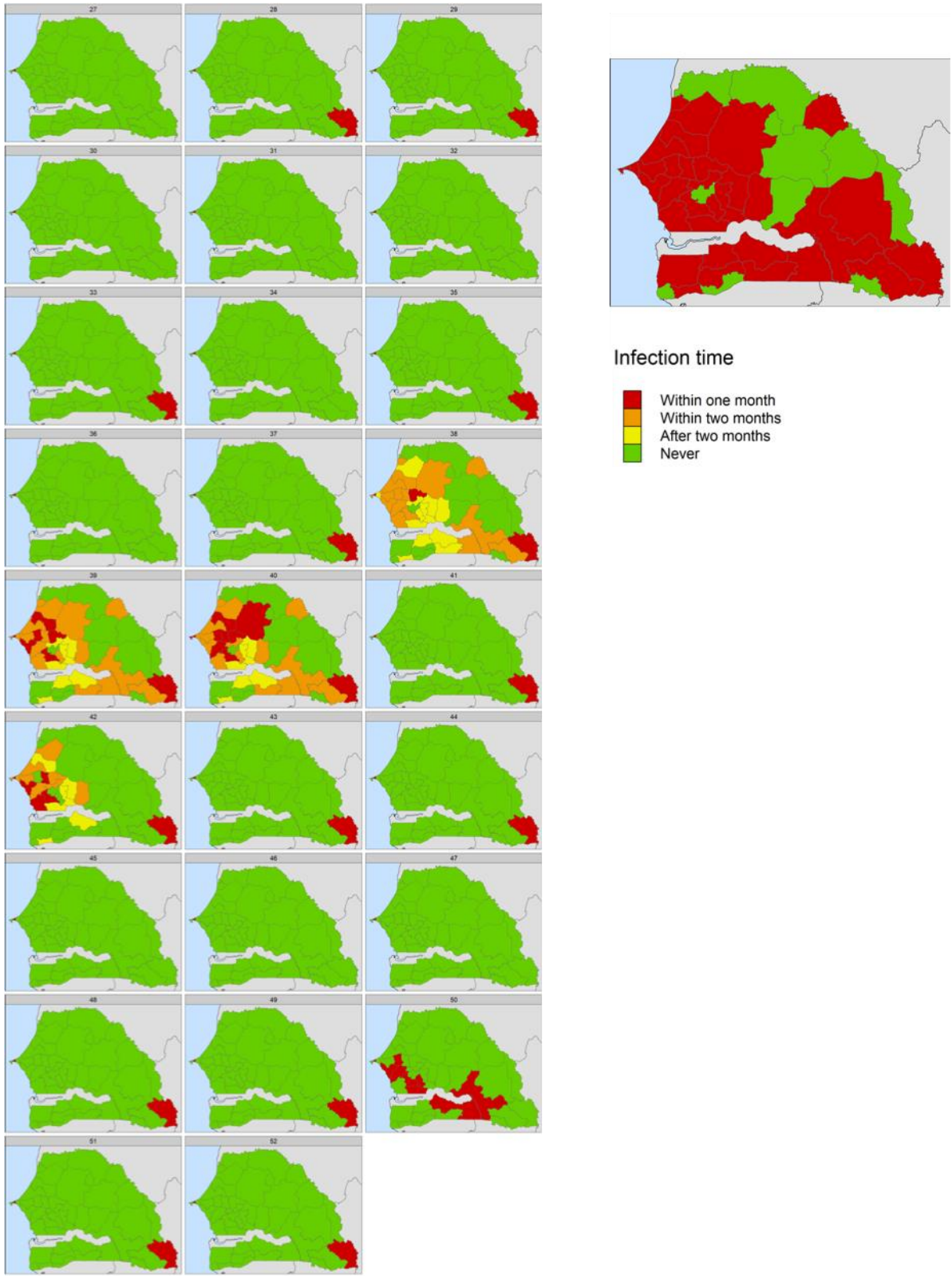

*SI Figure 7: Geographical representation of infection time in the case of a disease propagated from Mali through the cattle network. Temporal network on the left, static network on the right. For the static network, the colors are based on the links in the path: up to 5 in red, between 5 and 9 in orange, more than 9 in yellow. Nodes that have never been touched are colored green.*

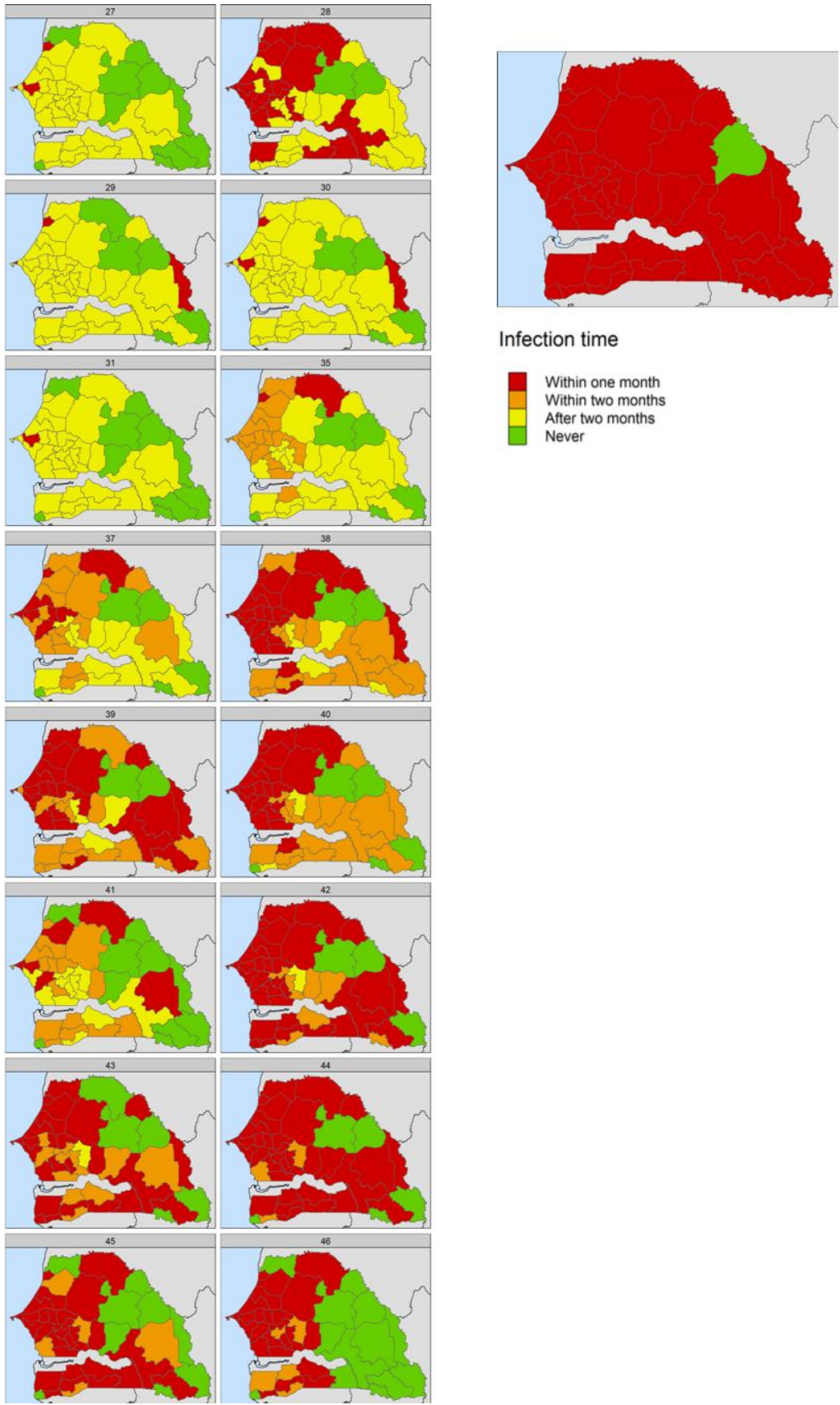

*SI Figure 8: Geographical representation of infection time in the case of a disease propagated from Mauritania through the livestock network. Temporal network on the left, static network on the right. For the static network, the colors are based on the links in the path: up to 5 in red, between 5 and 9 in orange, more than 9 in yellow. Nodes that have never been touched are colored green.*

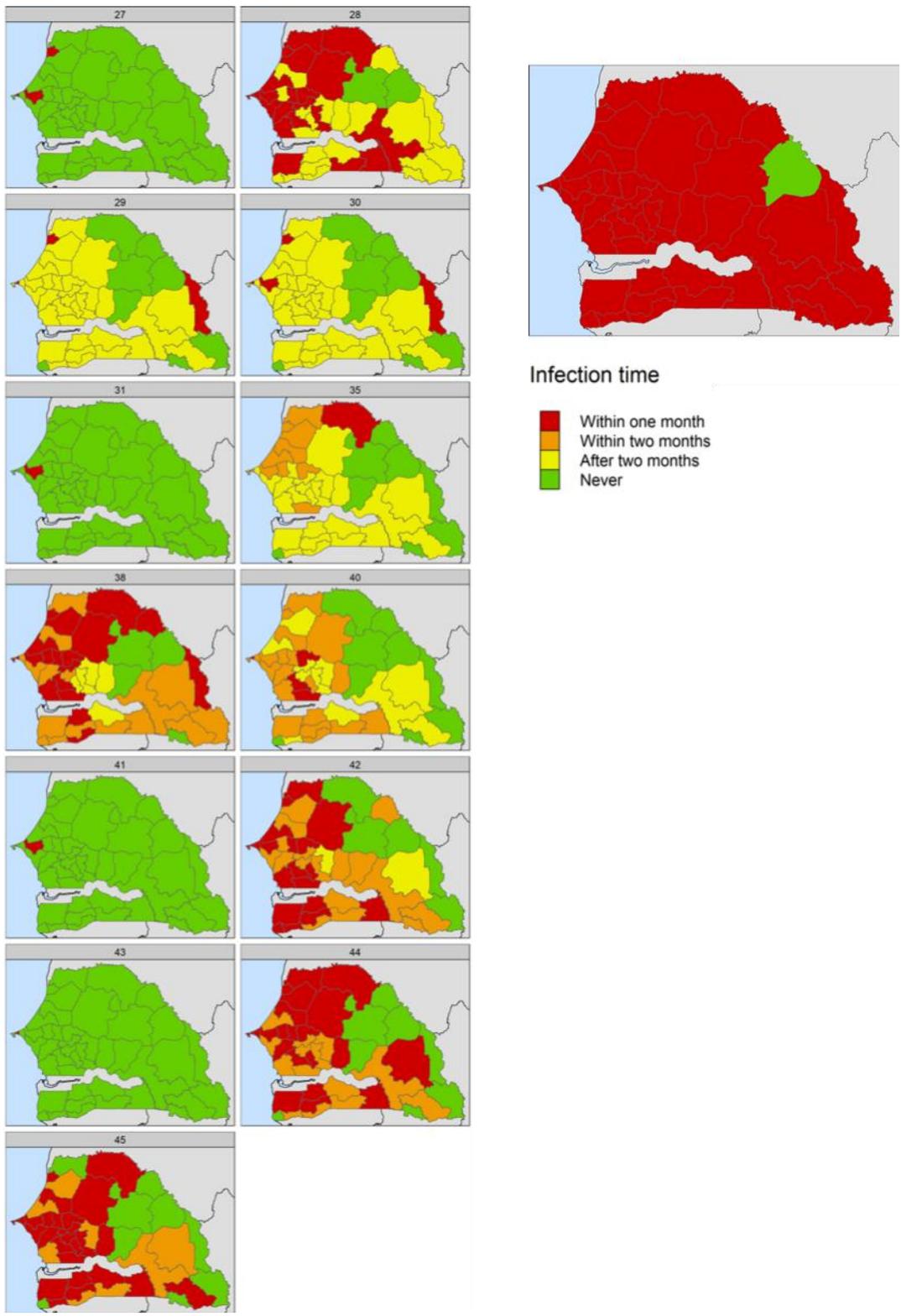

*SI Figure 9: Geographical representation of infection time in the case of a disease propagated from Mauritania through the small ruminant network. Temporal network on the left, static network on the right. For the static network, the colors are based on the links in the path: up to 5 in red, between 5 and 9 in orange, more than 9 in yellow. Nodes that have never been touched are colored green.*

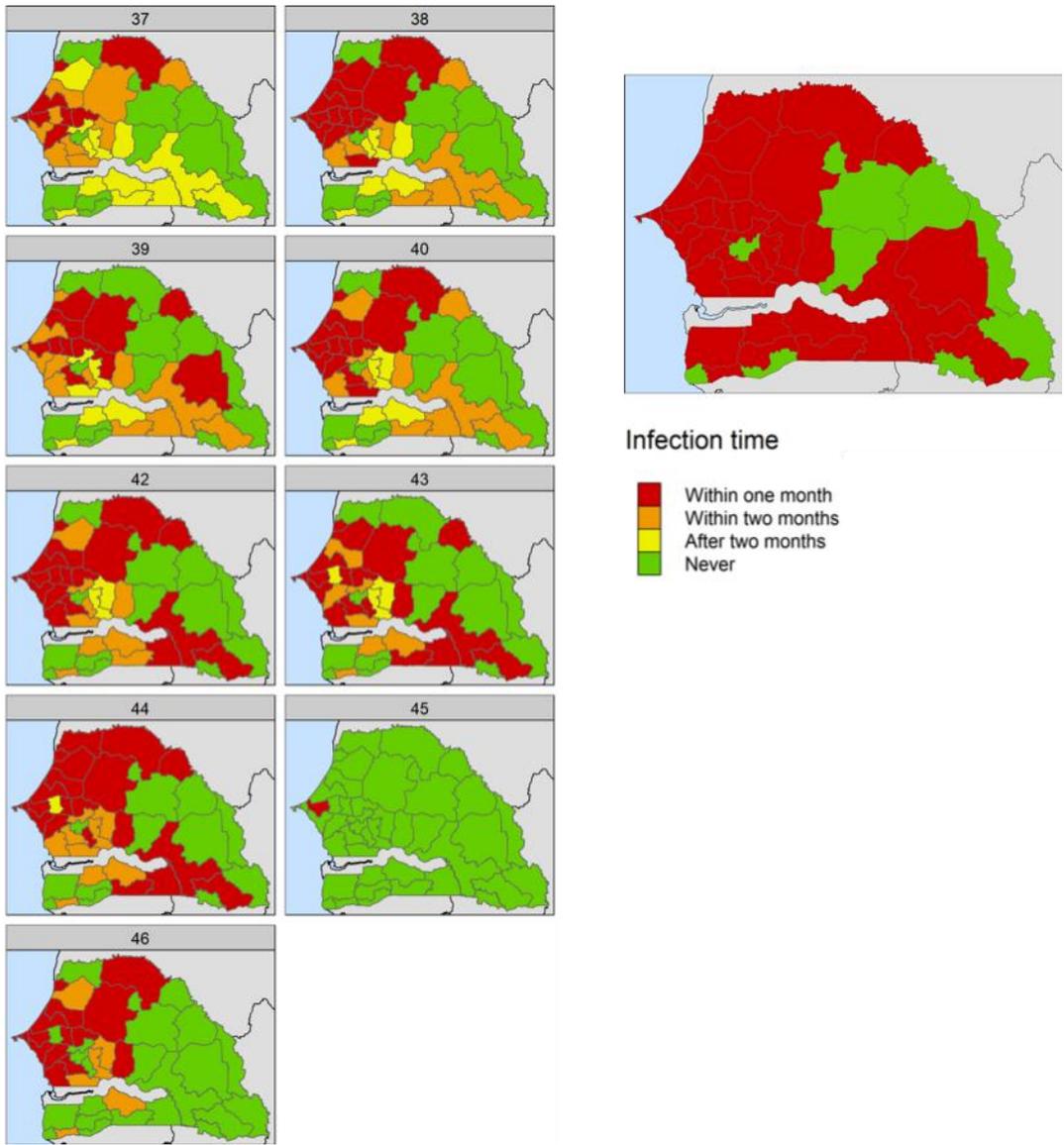

*SI Figure 10: Geographical representation of infection time in the case of a disease propagated from Mauritania through the cattle network. Temporal network on the left, static network on the right. For the static network, the colors are based on the links in the path: up to 5 in red, between 5 and 9 in orange, more than 9 in yellow. Nodes that have never been touched are colored green.*